\documentclass[conference,10pt,twoside]{IEEEtran}
\IEEEoverridecommandlockouts
\usepackage[T1]{fontenc}
\usepackage[utf8]{inputenc}
\usepackage{color}
\usepackage{mathrsfs}
\usepackage{bm}
\usepackage{amsmath}
\usepackage{amsthm}
\usepackage{amssymb}
\usepackage{cite}
\usepackage{balance}
\usepackage{subcaption}
\usepackage{algorithmic}
\usepackage[ruled,linesnumbered]{algorithm2e}

\usepackage{graphicx}
\usepackage{xcolor}
\usepackage{epstopdf}
\graphicspath{{./Figure/}}
\makeatletter
\theoremstyle{plain}
\newtheorem{thm}{\protect\theoremname}

\usepackage[caption=false,font=footnotesize]{subfig}

\usepackage{amsmath, amsfonts, amssymb, graphicx}

\newcommand{\herm}{^{\mathsf{H}}}
\newcommand{\trans}{^{\mathsf{T}}}

\DeclareMathOperator{\tr}{\mathsf{tr}}

\DeclareMathOperator{\diag}{\mathsf{diag}}
\DeclareMathOperator{\maximize}{maximize}
\DeclareMathOperator{\st}{subject~to}

\allowdisplaybreaks

\makeatother

\usepackage{babel}
\providecommand{\theoremname}{Theorem}
\begin{document}
\title{\huge{Achievable Rate Optimization for Large Flexible Intelligent Metasurface Assisted Downlink MISO under Statistical CSI}
}
\author{\IEEEauthorblockN{Ling~He\IEEEauthorrefmark{1}\IEEEauthorrefmark{2},
Vaibhav~Kumar\IEEEauthorrefmark{2}, 
Anastasios~Papazafeiropoulos\IEEEauthorrefmark{3}, 
Miaowen~Wen\IEEEauthorrefmark{1}\IEEEauthorrefmark{4},
Le-Nam~Tran\IEEEauthorrefmark{5},
and
Marwa Chafii\IEEEauthorrefmark{2}\IEEEauthorrefmark{6}}
\IEEEauthorblockA{\IEEEauthorrefmark{1}School of Electronic and Information Engineering, South China University of Technology, Guangzhou 510640, China}
\IEEEauthorblockA{\IEEEauthorrefmark{2}Wireless Research Lab, Engineering Division, New York University Abu Dhabi (NYUAD), UAE}
\IEEEauthorblockA{\IEEEauthorrefmark{3}Communications and Intelligent Systems Research Group, University of Hertfordshire, Hatfield AL10 9AB, U.K.}
\IEEEauthorblockA{\IEEEauthorrefmark{4}Guangdong Provincial Key Laboratory of Short-Range Wireless Detection and Communication, Guangzhou 510640, China}
\IEEEauthorblockA{\IEEEauthorrefmark{5}School of Electrical and Electronic Engineering, University College Dublin, Belfield, Dublin 4, Ireland}
\IEEEauthorblockA{\IEEEauthorrefmark{6}NYU WIRELESS, NYU Tandon School of Engineering, New York, USA}
Email: \{eelinghe@mail.scut.edu.cn, vaibhav.kumar@ieee.org, tapapazaf@gmail.com, \\
eemwwen@scut.edu.cn, nam.tran@ucd.ie, marwa.chafii@nyu.edu\}
\thanks{The work of Vaibhav Kumar and Marwa Chafii was supported by the Center for Cyber Security through New York University Abu Dhabi Research Institute under Award G1104. The work of Marwa Chafii was also supported by Tamkeen under the Research Institute NYUAD grant CG017.
The work of Ling He was supported by the China Scholarship Council (CSC) under Grant No. 202406150151, during her visiting Ph.D. study at NYUAD, UAE.
The work of Miaowen~Wen was supported by the Fundamental Research Funds for the Central Universities under Grant 2024ZYGXZR076.
The work of Le-Nam Tran was supported by Taighde Éireann – Research Ireland under Grant 22/US/3847 and Grant 13/RC/2077\_P2.}
}
\maketitle
\begin{abstract}
The integration of electromagnetic metasurfaces into wireless communications enables intelligent control of the propagation environment. Recently, flexible intelligent metasurfaces (FIMs) have evolved beyond conventional reconfigurable intelligent surfaces (RISs), enabling three-dimensional surface deformation for adaptive wave manipulation. However, most existing FIM-aided system designs assume perfect instantaneous channel state information (CSI), which is impractical in large-scale networks due to the high training overhead and complicated channel estimation. To overcome this limitation, we propose a robust statistical-CSI-based optimization framework for downlink multiple-input single-output (MISO) systems with FIM-assisted transmitters. A block coordinate ascent (BCA)-based iterative algorithm is developed to jointly optimize power allocation and FIM morphing, maximizing the average achievable sum rate. Simulation results show that the proposed statistical-CSI-driven FIM design significantly outperforms conventional rigid antenna arrays (RAAs), validating its effectiveness and practicality.
\end{abstract}


\IEEEpeerreviewmaketitle{}

\section{Introduction}\label{sec:Introduction}
With the rapid evolution toward sixth-generation (6G) networks, emerging applications such as industrial automation, autonomous driving, and the massive Internet of Things (IoT) impose stringent demands on electromagnetic (EM) wave control and environmental adaptability~\cite{24-COMST-6G-ISAC-Terahertz,23-COMST-6G}.
To address these challenges, recent research has shifted focus from traditional transceiver design to the intelligent reconfiguration of the wireless environment~\cite{25-OJCOMS-SurveyMagbool}.
Metamaterial-assisted transmitter and receiver architectures have therefore attracted considerable attention for their ability to manipulate EM waves using programmable subwavelength elements.
Among these, reconfigurable intelligent surfaces (RISs)~\cite{24-MVT-6G-RIS} have emerged as a leading implementation of metasurface technology, enabling low-cost and energy-efficient wireless channel reconfiguration through passive, programmable elements.
Nevertheless, the passive nature of RISs has hindered their practical deployment, owing to several challenges such as accurate channel estimation, control signaling overhead, and synchronization complexity.


To overcome these constraints, flexible intelligent metasurfaces (FIMs)~\cite{25-TAP-FIM} have emerged by integrating flexible electronics and smart materials, enabling three-dimensional deformable geometries that facilitate adaptive propagation control and enhanced diversity gains.
Recent studies have experimentally and numerically validated the potential of FIMs.
Specifically, An~\emph{et~al.}~\cite{25-TWC-FIM-MISO} demonstrated that an FIM-assisted multi-user multiple-input single-output (MISO) system can reduce the transmit power by at least 3~dB through joint beamforming and surface-shape optimization.
Their subsequent work~\cite{25-TCOMM-FIM-MIMO} extended this concept to multiple-input multiple-output (MIMO) systems, showing notable performance gains under weak channel conditions.
Furthermore, studies on transmissive-FIM and FIM-assisted multi-target wireless sensing have recently appeared in~\cite{25-TVT-FIM-relay} and~\cite{25-TVT-FIM-Sensing}, respectively.

However, existing works~\cite{25-TWC-FIM-MISO,25-TCOMM-FIM-MIMO,25-TVT-FIM-relay,25-TVT-FIM-Sensing} generally assume perfect instantaneous channel state information (CSI) and are restricted to the mmWave FR2 band (around 28~GHz), which limits their practicality in large-scale and low-frequency deployments.
In contrast, sub-6~GHz frequencies remain crucial for 6G networks, offering lower path loss, better coverage, and relaxed mechanical constraints for flexible deformation.
In this regime, FIMs can further exploit longer wavelengths to mitigate blockage, enhance spatial controllability, and improve mechanical robustness.
Meanwhile, obtaining perfect instantaneous CSI in large-scale FIM systems is unrealistic, as the channel coefficients are tightly coupled with the three-dimensional geometry, leading to excessive estimation overhead.

To overcome these challenges, we propose a robust statistical-CSI-based FIM design for downlink MISO systems operating in the sub-6 GHz band.
We first model the spatial correlation of the morphable elements and develop a minimum mean-square error (MMSE)-based channel estimation framework that yields a closed-form expression for the average achievable sum rate.
Then, a non-convex joint optimization of power allocation and surface morphing is formulated to maximize the average achievable sum rate. The problem is efficiently solved via a block coordinate ascent (BCA)-based iterative algorithm, which achieves robust performance without requiring instantaneous CSI.

\section{System Model and Problem Formulation}

\begin{figure}[t] 
    \centering
    \includegraphics[width=0.66\columnwidth]{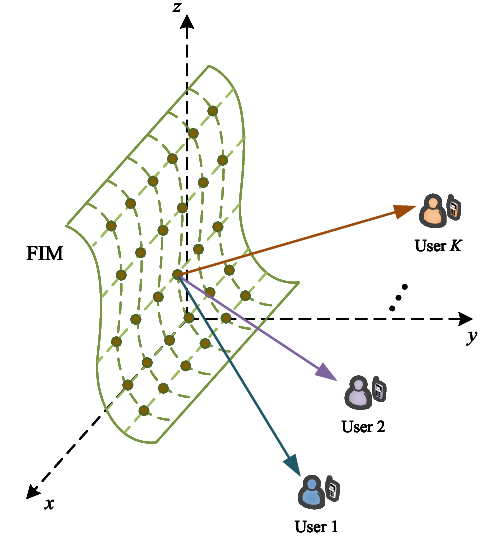}
    \caption{A large FIM-aided MISO system with multiple users.}
    \label{Fig_system}
\end{figure} 

We consider a multi-user MISO system, where a transmitter (Tx) simultaneously serves $K$ downlink users, as illustrated in Fig.~\ref{Fig_system}. 
The transmitter is equipped with a large FIM, modeled as a uniform planar array (UPA) located along the $x$--$z$ plane. 
The total number of transmit elements is denoted by $N = N_{x}N_{z}$, where $N_{x}$ and $N_{z}$ represent the numbers of transmit elements along the horizontal and vertical dimensions, respectively. 
Without loss of generality, we assume $N \gg K$. 
Under an isotropic scattering environment, the multipath components between the FIM and the users follow a uniform angular distribution expressed as $f(\vartheta,\varphi)=\cos(\vartheta)/2\pi,\vartheta\in[-\tfrac{\pi}{2},\tfrac{\pi}{2}],\varphi\in[-\pi,\pi]$, where $\vartheta$ and $\varphi$ denote the elevation and azimuth angles, respectively. 
Taking the transmit element at the origin as the reference, the location of the $n$-th transmit element of the FIM is given by $\mathbf{v}_{n}=[x_{n},y_{n},z_{n}]\trans\in\mathbb{R}^{3},\forall n\in\mathcal{N}\triangleq\{1,2,\ldots,N\}$, where $x_{n}=\mod{(n-1,N_{x})d_{\mathsf{H}}}$ and $z_{n}=\left\lfloor (n-1)/N_{x}\right\rfloor d_{\mathsf{V}}$. 
Here, $d_{\mathsf{H}}$ and $d_{\mathsf{V}}$ denote the horizontal and vertical inter-element spacings, respectively. 
The $y$-coordinate of the $n$-th transmit element satisfies the morphing constraint $y_{\min}\leq y_{n}\leq y_{\max},\forall n\in\mathcal{N}$, where $y_{\min}$ and $y_{\max}$ denote the minimum and maximum deformation limits along the $y$-axis. 
Following the convention~\cite{25-TWC-FIM-MISO}, we set $y_{\min}=0$, and define the morphed surface shape as $\mathbf{y}=[y_{1},\ldots,y_{N}]\trans\in\mathbb{R}^{N\times1}$. 
For a fixed $\mathbf{y}=\bm{0}$, the FIM reduces to the conventional rigid antenna array (RAA). 
Finally, we define the morphing range of the FIM as $r_{\mathrm{morph}}=y_{\max}-y_{\min}>0$, which characterizes the allowable level of surface deformation.

Let $L_{k}$ denote the number of distinct multipath components between the FIM and the $k$-th user $\mathrm{U}_{k},\forall k\in\mathcal{K}\triangleq\{1,2,\ldots,K\}$. 
The corresponding channel between the FIM and $\mathrm{U}_{k}$ is expressed as
\begin{equation}
\mathbf{h}_{k}(\mathbf{y})=\sum\nolimits_{l=1}^{L_{k}}\frac{c_{k,l}}{\sqrt{L_{k}}}\mathbf{a}(\mathbf{y},\vartheta_{l,k},\varphi_{l,k}),\label{eq:FIM-Uk-Channel}
\end{equation}
where $c_{k,l}/\sqrt{L_{k}}$ represents the complex-valued signal attenuation in the $l$-th path between the Tx and $\mathrm{U}_{k}$, $\mathbf{a}(\mathbf{y},\vartheta_{l,k},\varphi_{l,k})=[\exp(j\mathbf{w}(\vartheta_{l,k},\varphi_{l,k})\trans\mathbf{v}_{1}),\cdots,\exp(j\mathbf{w}(\vartheta_{l,k},\varphi_{l,k})\trans\mathbf{v}_{N})]\trans$ denotes the FIM response vector, $\mathbf{w}(\vartheta_{l,k},\varphi_{l,k})=\frac{2\pi}{\lambda}[\cos(\vartheta_{l,k})\cos(\varphi_{l,k}),\cos(\vartheta_{l,k})\sin(\varphi_{l,k}),\sin(\vartheta_{l,k})]\trans$, and $\lambda$ is the carrier wavelength. 
Assuming that $c_{k,l},\forall k,l$, are independent and identically distributed (i.i.d.) with zero mean and $A\mu_{k}$ variance, where $A$ denotes the area of a FIM transmitting element and $\mu_{k}$ the average intensity attenuation, the central-limit theorem yields $\lim\nolimits_{L_{k}\to\infty}\mathbf{h}_{k}(\mathbf{y})\sim\mathcal{CN}(\bm{0},\bm{\Sigma}_{k}(\mathbf{y}))$. 
Here, $\bm{\Sigma}_{k}(\mathbf{y})=A\mu_{k}\bm{\Sigma}_{\mathsf{FIM}}(\mathbf{y})$, and $\bm{\Sigma}_{\mathsf{FIM}}(\mathbf{y})\in\mathbb{R}^{N\times N}$ denotes the normalized FIM spatial correlation matrix, given by
\begin{equation}
\bm{\Sigma}_{\mathsf{FIM}}(\mathbf{y})=\frac{1}{A\mu_{k}}\mathbb{E}\{\mathbf{h}_{k}(\mathbf{y})\mathbf{h}_{k}\herm(\mathbf{y})\}.\label{eq:R_FIM-Matrix_Def}
\end{equation}
Following the arguments in~\cite[eqn.~(11)]{21-LWC-Emil-RayleighCorrelation} and after algebraic manipulation, the $(n,m)$-th element of $\bm{\Sigma}_{\mathsf{FIM}}(\mathbf{y})$ is obtained as
\begin{equation}
[\bm{\Sigma}_{\mathsf{FIM}}(\mathbf{y})]_{n,m}=\operatorname{sinc}\big(\tfrac{2\pi}{\lambda}\|\mathbf{v}_{n}-\mathbf{v}_{m}\|\big).\label{eq:R_FIM_element}
\end{equation}

\subsection{Channel Estimation}
Following the conventional time-division duplex (TDD) protocol in large MIMO systems, the channel estimation process is performed using uplink pilots. 
Let the training phase consist of $\tau$ pilot symbols. 
Each user $\mathrm{U}_{k}$ transmits an orthogonal pilot sequence $\tilde{\mathbf{x}}_{k}\in\mathbb{C}^{\tau\times1}$ satisfying $\tilde{\mathbf{x}}_{k}\herm\tilde{\mathbf{x}}_{k'}=0$ for $k\neq k'$ and $\|\tilde{\mathbf{x}}_{k}\|^{2}=\tau p_{\mathsf{train}}$, 
where $p_{\mathsf{train}}$ denotes the per-pilot-symbol transmit power. 
The received pilot matrix at the FIM is given by $\mathbf{Y}_{\mathsf{train}}(\mathbf{y})=\sum\nolimits_{k\in\mathcal{K}}\mathbf{h}_{k}(\mathbf{y})\,\tilde{\mathbf{x}}_{k}\herm+\mathbf{Z}_{\mathsf{train}}$, 
where $\mathbf{Z}_{\mathsf{train}}\in\mathbb{C}^{N\times\tau}$ has i.i.d. entries distributed as $\mathcal{CN}(0,\sigma^{2})$. 
Exploiting pilot orthogonality, the matched-filter statistic is obtained as $\mathbf{r}_{k}(\mathbf{y})=\mathbf{Y}_{\mathsf{train}}(\mathbf{y})\tilde{\mathbf{x}}_{k}/(\tau p_{\mathsf{train}})=\mathbf{h}_{k}(\mathbf{y})+\mathbf{z}_{\mathsf{train}}$, 
where $\mathbf{z}_{\mathsf{train}}\sim\mathcal{CN}(\bm{0},\tfrac{\sigma^{2}}{\tau p_{\mathsf{train}}}\mathbf{I})$. 
Given the zero-mean prior, the linear minimum mean-square error (LMMSE) estimate of $\mathbf{h}_{k}(\mathbf{y})$ is expressed as
\begin{equation}
\widehat{\mathbf{h}}_{k}(\mathbf{y})=\bm{\Sigma}_{k}(\mathbf{y})\mathbf{Q}_{k}(\mathbf{y})\mathbf{r}_{k}(\mathbf{y}),\label{eq:estimated_channel}
\end{equation}
where $\mathbf{Q}_{k}(\mathbf{y})=(\bm{\Sigma}_{k}(\mathbf{y})+\tfrac{\sigma^{2}}{\tau p_{\mathsf{train}}}\mathbf{I})^{-1}$. 
By the MMSE orthogonality principle~\cite{93-SMKay-StatisticalSignal}, the estimation error $\mathbf{e}_{k}(\mathbf{y})\triangleq\mathbf{h}_{k}(\mathbf{y})-\widehat{\mathbf{h}}_{k}(\mathbf{y})$ 
is uncorrelated with $\widehat{\mathbf{h}}_{k}(\mathbf{y})$, where
$\widehat{\bm{\Sigma}}_{k}(\mathbf{y})\triangleq\mathbb{E}\{\widehat{\mathbf{h}}_{k}(\mathbf{y})\widehat{\mathbf{h}}_{k}\herm(\mathbf{y})\}=\bm{\Sigma}_{k}(\mathbf{y})\mathbf{Q}_{k}(\mathbf{y})\bm{\Sigma}_{k}(\mathbf{y})$ 
and $\mathbf{E}_{k}(\mathbf{y})\triangleq\mathbb{E}\{\mathbf{e}_{k}(\mathbf{y})\mathbf{e}_{k}\herm(\mathbf{y})\}=\bm{\Sigma}_{k}(\mathbf{y})-\widehat{\bm{\Sigma}}_{k}(\mathbf{y})$.

\subsection{Downlink Data Transmission and Reception}
We consider a linear precoding scheme at the FIM Tx, where the transmitted signal is expressed as 
$\mathbf{x}=\mathbf{F}\mathbf{P}^{1/2}\mathbf{s}\in\mathbb{C}^{N\times1}$,
with $\mathbf{F}=[\mathbf{f}_{1},\ldots,\mathbf{f}_{K}]\in\mathbb{C}^{N\times K}$ 
denoting the linear precoding matrix employed to transmit the data symbol vector 
$\mathbf{s}=[s_{1},\ldots,s_{K}]\trans\sim\mathcal{CN}(\bm{0},\mathbf{I}_{K})\in\mathbb{C}^{K\times1}$ 
to $K$ downlink users. 
The diagonal matrix $\mathbf{P}=\diag(p_{1},\ldots,p_{K})\in\mathbb{R}^{K\times K}$ 
contains the allocated transmit powers $p_{k}\geq0$ for each user $\mathrm{U}_{k}$. 
The received signal at $\mathrm{U}_{k}$ is given by 
$y_{k}=\mathbf{h}_{k}\herm\mathbf{x}+z_{k}$, 
where $z_{k}\sim\mathcal{CN}(0,\sigma_{k}^{2})$ 
is the additive white Gaussian noise (AWGN) component. 

Since the transmission design in this work is optimized under statistical CSI, 
we invoke the well-established \emph{use-and-then-forget} (UaTF) bounding technique~\cite[Section~4.2]{17-bjornson-massiveMIMO}, 
which decomposes the effective channel into a deterministic mean gain and an uncorrelated random fluctuation. 
Accordingly, the received signal at $\mathrm{U}_{k}$ can be expressed as 
\begin{multline}
y_{k}=\underbrace{\sqrt{p_{k}}\mathbb{E}\{\mathbf{h}_{k}\herm\mathbf{f}_{k}\}s_{k}}_{\text{desired signal}}
+\underbrace{\big(\sqrt{p_{k}}\mathbf{h}_{k}\herm\mathbf{f}_{k}s_{k}-\sqrt{p_{k}}\mathbb{E}\{\mathbf{h}_{k}\herm\mathbf{f}_{k}\}s_{k}\big)}_{\text{beamforming uncertainty}}\\
+\underbrace{\sum\nolimits_{\jmath\in\mathcal{K}\setminus\{k\}}\sqrt{p_{\jmath}}\mathbf{h}_{k}\herm\mathbf{f}_{\jmath}s_{\jmath}}_{\text{multi-user interference}}+z_{k}.\label{eq:rx_signal_at_Uk}
\end{multline}

Let $\tau_{\mathsf{c}}$ denote the channel coherence interval (in symbols). 
Then, the achievable rate for $\mathrm{U}_{k}$ is given by\footnote{For analytical tractability, the achievable rate is expressed in nats/s/Hz; however, the simulation results are presented in bps/Hz.}  
\begin{equation}
R_{k}(\mathbf{F},\mathbf{p},\mathbf{y})=\bar{\tau}\ln(1+\gamma_{k}(\mathbf{F},\mathbf{P},\mathbf{y})),\label{eq:Rk}
\end{equation}
where $\bar{\tau}\triangleq\frac{\tau_{\mathsf{c}}-\tau}{\tau_{\mathsf{c}}}$ 
and $\mathbf{p}\triangleq[p_{1},\ldots,p_{K}]$ denotes the power allocation vector. 
The signal-to-interference-plus-noise ratio (SINR) at $\mathrm{U}_{k}$ is expressed as 
$\gamma_{k}(\mathbf{F},\mathbf{p},\mathbf{y})=S_{k}(\mathbf{F},\mathbf{p},\mathbf{y})/I_{k}(\mathbf{F},\mathbf{p},\mathbf{y})$, 
where
\begin{align}
S_{k}(\mathbf{F},\mathbf{p},\mathbf{y})=\ & p_{k}|\mathbb{E}\{\mathbf{h}_{k}\herm\mathbf{f}_{k}\}|^{2},\label{eq:useful_signal}\\
I_{k}(\mathbf{F},\mathbf{p},\mathbf{y})=\ & p_{k}\mathbb{E}\{|\mathbf{h}_{k}\herm\mathbf{f}_{k}-\mathbb{E}\{\mathbf{h}_{k}\herm\mathbf{f}_{k}\}|^{2}\}\nonumber\\
& +\sum\nolimits_{\jmath\in\mathcal{K}\setminus\{k\}}p_{\jmath}|\mathbb{E}\{\mathbf{h}_{k}\herm\mathbf{f}_{\jmath}\}|^{2}+\sigma_{k}^{2}.\label{eq:interference}
\end{align}

We remark that since the transmitter is equipped with a large FIM, 
computing the globally optimal precoding matrix $\mathbf{F}$ is computationally prohibitive. 
Therefore, we adopt the well-known \emph{maximal-ratio transmission} (MRT) scheme, 
which is widely employed in large MIMO systems owing to its near-optimal performance and low complexity. 
Under MRT, each precoding vector is aligned with its corresponding estimated channel, i.e., 
$\mathbf{f}_{k}=\widehat{\mathbf{h}}_{k},\forall k\in\mathcal{K}$, 
while the power allocation is optimized through $\mathbf{p}$. 

For the MRT case, with some algebraic manipulations, 
the closed-form expressions of~\eqref{eq:useful_signal} and~\eqref{eq:interference} can be obtained as\footnote{Detailed derivations are omitted due to space constraints.}
\begin{align}
S_{k}(\mathbf{p},\mathbf{y})=\ & p_{k}(\tr(\widehat{\bm{\Sigma}}_{k}(\mathbf{y})))^{2},\label{eq:closed_S}\\
I_{k}(\mathbf{p},\mathbf{y})=\ & \sum\nolimits_{\jmath\in\mathcal{K}}p_{\jmath}\tr(\bm{\Sigma}_{k}(\mathbf{y})\widehat{\bm{\Sigma}}_{\jmath}(\mathbf{y}))\nonumber\\
& \qquad\qquad-p_{k}\tr(\widehat{\bm{\Sigma}}_{k}^{2}(\mathbf{y}))+\sigma_{k}^{2}.\label{eq:closed_I}
\end{align}
Note that in~\eqref{eq:closed_S} and~\eqref{eq:closed_I}, 
$\mathbf{F}$ has been omitted as a variable since the precoder is fixed to MRT. 
Consequently, the SINR now depends solely on the power allocation $\mathbf{p}$ 
and the surface morphing configuration $\mathbf{y}$, 
and the achievable rate for $\mathrm{U}_{k}$ can be compactly expressed as $R_{k}(\mathbf{p},\mathbf{y})$.

\subsection{Problem Formulation}
With the above background, we now formulate the joint optimization problem of determining the optimal power allocation $\mathbf{p}$ and FIM morphing configuration $\mathbf{y}$ that maximize the achievable sum rate, subject to individual rate constraints, total transmit power budget, and morphing range limitations. 
Recalling the transmitted signal 
$\mathbf{x}=\mathbf{F}\mathbf{P}^{1/2}\mathbf{s}=\sum\nolimits_{k\in\mathcal{K}}\sqrt{p_{k}}\mathbf{f}_{k}s_{k}$, 
and adopting MRT-based precoding, the average (long-term) transmit power is given by 
\begin{equation}
\!\!\mathbb{E}\{\|\mathbf{x}\|^{2}\}
\!=\!\sum\nolimits_{k\in\mathcal{K}}p_{k}\mathbb{E}\{\|\widehat{\mathbf{h}}_{k}\|^{2}\}
\!=\!\sum\nolimits_{k\in\mathcal{K}}p_{k}\tr(\widehat{\bm{\Sigma}}_{k}(\mathbf{y})).\label{eq:total_transmit_power}\!\!
\end{equation}
Accordingly, the sum-rate maximization problem for the FIM-assisted downlink system can be formulated as 
\begin{subequations}
\label{eq:P1}
\begin{align}
\underset{\mathbf{p},\mathbf{y}}{\maximize}\ \ & \sum\nolimits_{k\in\mathcal{K}}R_{k}(\mathbf{p},\mathbf{y})\label{eq:P1-obj}\\
\st\ \ & R_{k}(\mathbf{p},\mathbf{y})\geq r_{k,\min},\ \forall k\in\mathcal{K},\label{eq:P1-QoS}\\
& \sum\nolimits_{k\in\mathcal{K}}p_{k}\tr(\widehat{\bm{\Sigma}}_{k}(\mathbf{y}))\leq p_{\max},\label{eq:P1-TPC}\\
& 0\leq y_{n}\leq y_{\max},\ \forall n\in\mathcal{N},\label{eq:P1-MoC}
\end{align}
\end{subequations}
where~\eqref{eq:P1-obj} represents the achievable sum rate, \eqref{eq:P1-QoS} guarantees that each user $\mathrm{U}_{k}$ attains its minimum rate threshold $r_{k,\min}$, \eqref{eq:P1-TPC} enforces the total transmit power constraint $p_{\max}$, and \eqref{eq:P1-MoC} restricts the morphing displacement of each FIM element within the allowable range.

The optimization problem in~\eqref{eq:P1} is non-convex due to the coupled variables $\mathbf{p}$ and $\mathbf{y}$ and the non-linear structure of $R_{k}(\mathbf{p},\mathbf{y})$, rendering it analytically intractable. 
To obtain a high-quality stationary solution, we adopt a \emph{block coordinate ascent} (BCA) framework, in which~\eqref{eq:P1} is decomposed into two tractable subproblems: 
(i) the power allocation optimization, and 
(ii) the FIM morphing optimization. 
Each subproblem is solved by optimizing one block of variables while keeping the other fixed, and the procedure iteratively alternates between the two until convergence to a stationary point is achieved. 
The proposed solution methodology is detailed in the following section.

\section{BCA-Based Proposed Solution}

To efficiently solve the non-convex optimization problem in~(\ref{eq:P1}), 
we employ a BCA framework, 
where the coupled optimization variables are updated alternately. 
In each iteration, one variable block (either power allocation or FIM morphing) is optimized while keeping the other fixed, 
ensuring a non-decreasing objective sequence that converges to a stationary point.

\subsection{Optimal Power Allocation for a Fixed Morphing}

For a given FIM morphing configuration $\mathbf{y}$, the optimization problem in~\eqref{eq:P1} reduces to
\begin{subequations}
\label{eq:P2}
\begin{align}
\underset{\mathbf{p}}{\maximize}\ \ & \sum\nolimits_{k\in\mathcal{K}}R_{k}(\mathbf{p})\label{eq:P2-obj}\\
\st\ \ & R_{k}(\mathbf{p})\geq r_{k,\min},\ \forall k\in\mathcal{K},\label{eq:P2-QoS}\\
& \sum\nolimits_{k\in\mathcal{K}}p_{k}\tr(\widehat{\bm{\Sigma}}_{k})\leq p_{\max}.\label{eq:P2-TPC}
\end{align}
\end{subequations}
Here, the dependence on $\mathbf{y}$ is omitted since the FIM geometry is fixed during this step. 
Problem~\eqref{eq:P2} remains \emph{non-convex}, as each $R_{k}(\mathbf{p})=\bar{\tau}\ln(1+\gamma_{k}(\mathbf{p}))$ involves a logarithmic function composed with a fractional SINR term, 
where both the numerator and denominator are affine functions of $\mathbf{p}$. 
This structure renders $R_{k}(\mathbf{p})$ neither concave nor convex, 
making direct optimization intractable.

To address this challenge, we employ the \emph{successive convex approximation} (SCA) technique, 
which iteratively constructs a locally tight concave lower bound of the non-convex objective around the current feasible point $\widehat{\mathbf{p}}^{(\varkappa)}$ at iteration $\varkappa$. 
Specifically, the term $\ln(1+\gamma_{k}(\mathbf{p}))$ is approximated by its first-order Taylor expansion at $\widehat{\mathbf{p}}^{(\varkappa)}$, ensuring a convex surrogate subproblem that preserves global feasibility while improving the objective value. 
The resulting convex approximation of~\eqref{eq:P2} is characterized in the following theorem.

\begin{thm}
\label{thm:convex_reform}
At iteration $\varkappa$, a convex surrogate of~\eqref{eq:P2} can be formulated as
\begin{subequations}
\label{eq:P4}
\begin{align}
\underset{\widehat{\mathbf{p}}}{\maximize}\ \ & \sum\nolimits_{k\in\mathcal{K}}R_{k}(\widehat{\mathbf{p}};\widehat{\mathbf{p}}^{(\varkappa)})\label{eq:P4-obj}\\
\st\ \ & R_{k}(\widehat{\mathbf{p}};\widehat{\mathbf{p}}^{(\varkappa)})\geq r_{k,\min},\ \forall k\in\mathcal{K},\label{eq:P4-QoS}\\
& \sum\nolimits_{k\in\mathcal{K}}\widehat{p}_{k}\leq p_{\max},\label{eq:P4-TPC}
\end{align}
\end{subequations}
where $R_{k}(\widehat{\mathbf{p}};\widehat{\mathbf{p}}^{(\varkappa)})$ denotes the concave lower-bound approximation of $R_{k}(\mathbf{p})$ around the current iterate $\widehat{\mathbf{p}}^{(\varkappa)}$.
\end{thm}

\begin{IEEEproof}
See Appendix~\ref{sec:convex_reform_proof}.
\end{IEEEproof}

The surrogate subproblem~\eqref{eq:P4} is a standard convex program, 
characterized by a concave objective and convex constraints. 
It can thus be efficiently solved using off-the-shelf packages, e.g., CVX/CVXPY. 
The optimal solution is then used as the updated iterate for the next BCA iteration, 
ensuring monotonic non-decreasing improvement of the overall sum-rate objective until convergence to a stationary point.

\subsection{Optimal Morphing for a Fixed Power Allocation}

For fixed power allocation $\mathbf{p}$,~\eqref{eq:P1} reduces to
\begin{subequations}
\label{eq:P5}
\begin{align}
\underset{\mathbf{y}}{\maximize}\ \ & \sum\nolimits_{k\in\mathcal{K}}R_{k}(\mathbf{y})\label{eq:P5-obj}\\
\st\ \ & R_{k}(\mathbf{y})\geq r_{k,\min},~\forall k,\label{eq:P5-QoS}\\
& \sum\nolimits_{k\in\mathcal{K}}p_{k}\tr(\widehat{\bm{\Sigma}}_{k}(\mathbf{y}))\leq p_{\max},\label{eq:P5-TPC}\\
& \eqref{eq:P1-MoC}.\nonumber
\end{align}
\end{subequations}
Since $\mathbf{p}$ is fixed,~\eqref{eq:P5} depends solely on the morphing vector $\mathbf{y}$. 
However, $R_{k}(\mathbf{y})$ is highly non-convex in $\mathbf{y}$ due to the nonlinear coupling in $\widehat{\bm{\Sigma}}_{k}(\mathbf{y})$. 
To efficiently handle the coupled non-convex constraints, we blend penalty and dual decomposition principles. For this purpose, we define 
\begin{equation}
g_{k}(\mathbf{y},r_{k,\min},\varepsilon_{k})\triangleq r_{k,\min}+\varepsilon_{k}-R_{k}(\mathbf{y}),\forall k\in\mathcal{K},\label{eq:gk}
\end{equation}
\begin{equation}
g_{0}(\mathbf{y},p_{\max},\varepsilon_{0})\triangleq\frac{1}{p_{\max}}\sum\nolimits_{k\in\mathcal{K}}p_{k}\tr(\widehat{\mathbf{\bm{\Sigma}}}_{k}(\mathbf{y}))+\varepsilon_{0}-1,\label{eq:g0}
\end{equation}
where $\bm{\varepsilon}\!\triangleq\![\varepsilon_{0},\ldots,\varepsilon_{K}]\trans\!\in\!\mathbb{R}_{\ge0}^{K+1}$ are slack variables ensuring feasibility. 
The augmented Lagrangian associated with~\eqref{eq:P5} is then given by
\begin{align}
\!\! & \mathscr{F}_{\bm{\nu},\rho}(\mathbf{y},\bm{\varepsilon})\nonumber \\
\!\!= & \sum\nolimits_{k\in\mathcal{K}}\Big[R_{k}(\mathbf{y})\!-\!\tau_{k}g_{k}(\mathbf{y},r_{k,\min},\varepsilon_{k})\Big]\!-\!\tau_{0}g_{0}(\mathbf{y},p_{\max},\varepsilon_{0})\nonumber \\
\!\! & -\frac{1}{2\rho}\Big[\sum\nolimits_{k\in\mathcal{K}}g_{k}^{2}(\mathbf{y},r_{k,\min},\varepsilon_{k})+g_{0}^{2}(\mathbf{y},p_{\max},\varepsilon_{0})\Big],\label{eq:augObj}
\end{align}
where $\bm{\nu}\!\triangleq\![\nu_{0},\ldots,\nu_{K}]\trans$ is the vector of Lagrange multipliers and $\rho>0$ is the penalty parameter. 
The corresponding subproblem is formulated as
\begin{equation}
\underset{\mathbf{y},\,\bm{\varepsilon}\ge0}{\maximize}\ 
\{\mathscr{F}_{\bm{\nu},\rho}(\mathbf{y},\bm{\varepsilon})\ \big|\ \eqref{eq:P1-MoC}\}.
\label{eq:P6}
\end{equation}
Since~\eqref{eq:P6} is still non-convex, a gradient-ascent strategy is adopted to reach a stationary solution. 
The closed-form gradient of $\mathscr{F}_{\bm{\nu},\rho}(\mathbf{y},\bm{\varepsilon})$ with respect to $y_{n}$ is derived below.

\begin{thm}
\label{thm:grad_theorem}A closed-form expression for $\nabla_{y_{n}}\mathscr{F}_{\bm{\nu},\rho}(\mathbf{y},\bm{\tau})$
is given by~(\ref{eq:grad_closed}), shown on the next page. 
\begin{figure*}
\begin{equation}
\nabla_{y_{n}}\mathscr{F}_{\bm{\nu},\rho}(\mathbf{y},\bm{\tau})=\sum_{k\in\mathcal{K}}\Big[\big\{1+\nu_{k}+\tfrac{1}{\rho}g_{k}(\mathbf{y},r_{k,\min},\varepsilon_{k})\big\}\nabla_{y_{n}}R_{k}(\mathbf{y})\Big]-\frac{1}{p_{\max}}\big\{\nu_{0}+\tfrac{1}{\rho}g_{0}(\mathbf{y},p_{\max},\varepsilon_{0})\big\}\sum_{k\in\mathcal{K}}p_{k}\nabla_{y_{n}}\tr(\widehat{\bm{\Sigma}}_{k}(\mathbf{y})).\label{eq:grad_closed}
\end{equation}

\rule{1\textwidth}{1pt}
\end{figure*}
\end{thm}

\begin{IEEEproof}
See Appendix~\ref{sec:grad_proof}.
\end{IEEEproof}

Finally, the proposed BCA-based algorithm for optimizing $\mathbf{p}$ and $\mathbf{y}$ is summarized in \textbf{Algorithm~\ref{algorithm-1}}, where $\mathbf x^{(\jmath)}$ denotes the value of the optimization variable $\mathbf x$ in the $\jmath$-th iteration, $\delta$ is the step-size obtained using the Barzilai and Borwein method, and $\Pi_{\mathcal Y_n}\{x\} \triangleq \max \{\min\{x, y_{\max}\}, 0\}$  is the projection operator onto the feasible set of $y_n$, according to the maximum morphing range $y_{\max}$. 
Furthermore, the per-iteration complexity of \textbf{Algorithm~\ref{algorithm-1}} is mainly determined by lines 4 and 9, with respective complexities of $\mathcal{O}(K^{3})$~\cite{11-BenTal-Convex-Opt} and $\mathcal{O}(N^3)$.
Since $N \gg K$, the overall complexity of \textbf{Algorithm~\ref{algorithm-1}} can be approximated as $\mathcal{O}(N^{3})$.

\begin{algorithm}[tb]
\caption{The Proposed BCA-Based Algorithm.}

\label{algorithm-1}

\KwIn{$\mathbf{p}^{(0)}$, $\mathbf y^{(0)}$, $\bm{\varepsilon}^{(0)}$,
$\bm{\nu}$, $\rho$, $\varsigma$}

\KwOut{$\mathbf{p}^{\mathrm{opt}}$, $\mathbf y^{\mathrm{opt}}$}

\Repeat{convergence }{


$\jmath\leftarrow1$\;

\Repeat{convergence}{
Obtain $\mathbf{p}^{(\jmath)}$ by solving~\eqref{eq:P2}\;
$\jmath \leftarrow \jmath + 1$\;
}

\Repeat{convergence }{


\Repeat{convergence}{
$\mathbf{y}^{(\jmath)}_n\!\leftarrow\!\Pi_{\mathcal{Y}_n}\big\{\!\mathbf{y}^{(\jmath\!-\!1)}_n\!+\!\delta\nabla_{y_n}\mathscr{F}_{\bm{\nu},\rho}\big(\mathbf{y}^{(\jmath\!-\!1)},\boldsymbol{\varepsilon}^{(\jmath\!-\!1)}\big)\!\big\},\forall n \in \mathcal N$\;

$\varepsilon_0^{(\jmath)} = \max\{0, 1 \!-\! \frac{1}{p_{\max}}\!\!\sum_{k \in \mathcal K} p_k \tr (\widehat{\bm{\Sigma}}_k (\mathbf y^{(\jmath - 1)})) \!-\! \nu_0 \rho\}$\;

$\varepsilon_{k}^{(\jmath)}=\max\{0,R_k(\mathbf y^{(\jmath-1)}) - r_{k,\min}-\nu_{k}\rho\},\forall k\in\mathcal{K}$\;


$\jmath \leftarrow \jmath + 1$\;
}
$\nu_0 \leftarrow \nu_0 + \frac{1}{\rho} g_0(\mathbf y^{(\jmath)}, p_{\max}, \varepsilon_0)$\;
$\nu_k \leftarrow \nu_k + \frac{1}{\rho} g_k(\mathbf y^{(\jmath)}, r_{k, \min}, \varepsilon_k),\forall k \in \mathcal K$\;
$\rho \leftarrow \varsigma \rho$\; 
} 
}
$\mathbf{p}^{\mathrm{opt}} \leftarrow \mathbf{p}^{\jmath}$, $\mathbf y^{\mathrm{opt}} \leftarrow \mathbf y^{(\jmath)}$
\end{algorithm}

\begin{figure*}
\centering
\begin{minipage}{.32\textwidth}
  \centering
  \includegraphics[width=1\columnwidth]{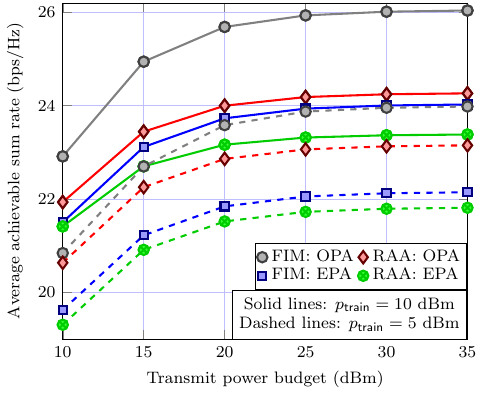}		
  \caption{Impact of the transmit power budget on the average achievable sum rate.}
  \label{Fig_Power}
\end{minipage}%
\hfill 
\begin{minipage}{.32\textwidth}
  \centering
  \includegraphics[width=1\columnwidth]{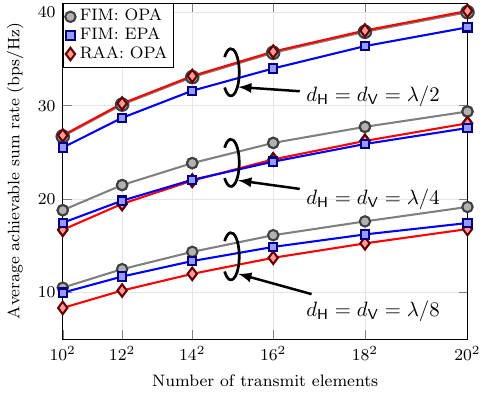}		
  \caption{Impact of the number of transmit elements on the average achievable sum rate.}
	\label{Fig_elements}
\end{minipage}%
\hfill 
\begin{minipage}{.32\textwidth}
  \centering
  \includegraphics[width=1\columnwidth]{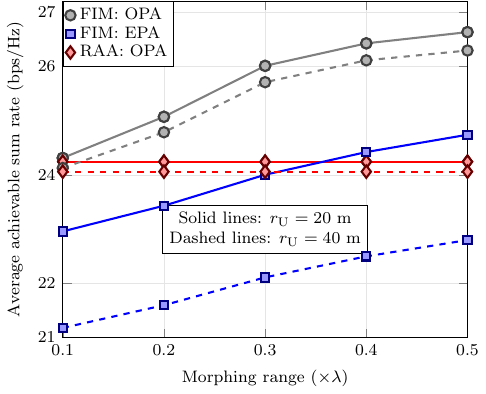}		
	\caption{Impact of the morphing range on the average achievable sum rate.}
	\label{Fig_Morphing}
\end{minipage} 
\end{figure*}

\section{Simulation Results}
In the considered FIM-assisted downlink system, the reference element of the FIM is positioned at $(0,0,0)$~m, while the users are uniformly distributed within a circular area of radius $r_{\mathrm{U}} = 20$~m, centered $50$~m away from the reference element. 
The system operates at a carrier frequency of $3.5$~GHz with a $20$~MHz bandwidth, and the noise power spectral density at each user is set to $-174$~dBm/Hz. 
Unless otherwise specified, the default parameters are $N = 16^{2}$, $K = 8$, $\tau_{\mathsf{c}} = 200$, $\tau = K$, $r_{k,\min}=r_{\min}=1~\mathrm{bps/Hz},\,\forall k\in\mathcal{K}$, $p_{\max}=30$~dBm, $p_{\mathsf{train}}=10$~dBm, $d_{\mathsf{H}}=d_{\mathsf{V}}=\lambda/4$, and $y_{\max}=0.3\lambda$. 
The large-scale path loss between the FIM and user $\mathrm{U}_{k}$ follows $(-30 - 10\wp\log_{10}(d_{k}/d_{0}))$~dB, where $\wp=2.8$ is the path-loss exponent, $d_{k}$ is the distance between the FIM and $\mathrm{U}_{k}$, and $d_{0}=1$~m is the reference distance. 
Each reported result is obtained by averaging over $100$ random user realizations. 
For benchmarking, OPA refers to the optimal per-user power allocation obtained from the proposed algorithm, while EPA denotes the equal-power allocation baseline.

Figure~\ref{Fig_Power} illustrates the variation of the average achievable sum rate with the transmit power budget $p_{\max}$. 
As expected, the sum rate increases with $p_{\max}$ and saturates beyond $25$~dBm, marking the transition from the noise-limited to the interference-limited regime. 
The proposed FIM-assisted design with optimal power allocation (OPA) consistently achieves the highest performance, whereas the RAA with equal power allocation (EPA) yields the lowest. 
For $p_{\mathsf{train}} = 10$~dBm, achieving an average sum rate of approximately $23$~bps/Hz requires about 5~dB less transmit power under FIM–OPA than under FIM–EPA, and nearly 10~dB less than the RAA–EPA scheme. 
This confirms that joint optimization of FIM morphing and transmit power greatly enhances power efficiency by maintaining the same throughput with substantially lower transmit power. 
Moreover, OPA-based schemes outperform their EPA counterparts for both FIM and RAA configurations, highlighting the robustness of the proposed power allocation method. 
Finally, increasing $p_{\mathsf{train}}$ from $5$ to $10$~dBm improves channel estimation accuracy, thereby raising the achievable sum rate, yielding a $2.06$~bps/Hz gain for FIM compared to $1.11$~bps/Hz for RAA, under the OPA scheme, which further underscores the adaptability of FIM.

In Fig.~\ref{Fig_elements}, we examine the impact of the number of transmit elements $N$ and the inter-element spacing $d_{\mathsf{E}} = d_{\mathsf{H}} = d_{\mathsf{V}}$ on the average achievable sum rate. 
As expected, the sum rate increases monotonically with $N$, and the proposed FIM with OPA consistently achieves superior performance compared to other schemes. 
Under OPA, the performance gap between the FIM and RAA is most pronounced at $d_{\mathsf{E}} = \lambda/8$, but gradually diminishes with increasing $d_{\mathsf{E}}$. 
At $d_{\mathsf{E}} = \lambda/2$, the two curves nearly coincide, as reduced spatial correlation and limited morphing flexibility constrain the FIM advantage. 
Moreover, when $d_{\mathsf{E}} = \lambda/8$, increasing $N$ amplifies the rate gain of OPA over EPA for the FIM, highlighting the benefit of larger spatial degrees of freedom for power adaptation. 
Interestingly, the gap between FIM–EPA and RAA–OPA narrows with $N$, indicating that optimized power allocation partially compensates for the geometric rigidity of RAA. 
For aperture-limited deployments, smaller $d_{\mathsf{E}}$ values enable denser element placement and higher spatial resolution; for instance, in the FIM–OPA case, reducing $d_{\mathsf{E}}$ from $\lambda/2$ to $\lambda/4$ increases the number of deployable elements from $10^{2}$ to (nearly) $20^{2}$, improving the average sum rate from $26.71$ to $29.38$~bps/Hz.

Finally, Fig.~\ref{Fig_Morphing} illustrates the impact of the morphing range $r_{\mathrm{morph}}$ on the average achievable sum rate. 
The performance of the RAA remains constant due to its fixed geometry, whereas the FIM achieves steadily higher rates as $r_{\mathrm{morph}}$ increases, reflecting the advantage of enhanced spatial flexibility. 
Under OPA with $r_{\mathrm{U}}=20$~m, enlarging $r_{\mathrm{morph}}$ from $0.1\lambda$ to $0.5\lambda$ improves the FIM sum rate from $24.31$ to $26.63$~bps/Hz, corresponding to a $2.39$~bps/Hz gain. 
Moreover, the rate gap between OPA and EPA for FIM widens from $1.89$~bps/Hz at $r_{\mathrm{U}}=20$~m to $3.50$~bps/Hz at $r_{\mathrm{U}}=40$~m, indicating that a larger user area offers more flexibility for adaptive power allocation. 
Although increasing $r_{\mathrm{U}}$ slightly degrades overall performance due to greater path loss, the OPA schemes exhibit significantly smaller degradation than EPA, confirming that optimized power allocation effectively mitigates the impact of unfavorable user distributions.

\section{Conclusion}
This paper investigated a statistical-CSI-based FIM-assisted downlink MISO system in the sub-6~GHz band. We derived the spatial correlation model for the morphable FIM and proposed an MMSE-based channel estimation approach, leading to a closed-form rate expression. A BCA-based algorithm was then developed to jointly optimize the power allocation and surface morphing. Simulation results confirmed that the proposed design consistently outperforms conventional RAA and EPA schemes, with clear gains under diverse user distributions and morphing configurations. These findings highlight the potential of FIMs to realize adaptive and high-throughput reconfigurable wireless systems.

\appendices{}

\section{Proof of Theorem~\ref{thm:convex_reform}}\label{sec:convex_reform_proof}

Let $\widehat{p}_{k}\triangleq p_{k}\tr(\widehat{\bm{\Sigma}}_{k})$ and 
$\widehat{\mathbf{p}}\triangleq[\widehat{p}_{1},\ldots,\widehat{p}_{K}]\trans\!\in\!\mathbb{R}^{K\times1}$. 
Then,~\eqref{eq:P2} can be equivalently rewritten as
\begin{subequations}
\label{eq:P3}
\begin{align}
\underset{\widehat{\mathbf{p}}}{\maximize}\ \ & \sum\nolimits_{k\in\mathcal{K}}R_{k}(\widehat{\mathbf{p}})\label{eq:P3-obj}\\
\st\ \ & R_{k}(\widehat{\mathbf{p}})\geq r_{k,\min},~\forall k,\label{eq:P3-QoS}\\
& \sum\nolimits_{k\in\mathcal{K}}\widehat{p}_{k}\leq P_{\max}.\label{eq:P3-TPC}
\end{align}
\end{subequations}
Both~\eqref{eq:P3-obj} and~\eqref{eq:P3-QoS} are non-convex due to the fractional SINR term in the achievable rate.  
By defining $\gamma_{k}(\widehat{\mathbf{p}})=
\frac{\psi_{k}\widehat{p}_{k}}
{\bar{\bm{\psi}}_{k}\trans\widehat{\mathbf{p}}+\sigma_{k}^{2}}$, 
where $\psi_{k}=\tr(\widehat{\bm{\Sigma}}_{k})$ and 
$\bar{\bm{\psi}}_{k}=[\bar{\psi}_{k,1},\ldots,\bar{\psi}_{k,K}]\trans$ with
\begin{equation}
\bar{\psi}_{k,\imath}=
\begin{cases}
\dfrac{\tr(\bm{\Sigma}_{k}\widehat{\bm{\Sigma}}_{\imath})}{\tr(\widehat{\bm{\Sigma}}_{\imath})}, & \text{if }\imath\neq k,\\[6pt]
\dfrac{\tr(\bm{\Sigma}_{k}\widehat{\bm{\Sigma}}_{k})}{\tr(\widehat{\bm{\Sigma}}_{k})}
-\dfrac{\tr(\widehat{\bm{\Sigma}}_{k}^{2})}{\tr(\widehat{\bm{\Sigma}}_{k})}, & \text{if }\imath=k,
\end{cases}
\label{eq:bar_psi}
\end{equation}
the achievable rate can be expressed as
\begin{equation}
R_{k}(\widehat{\mathbf{p}})=
\ln(\psi_{k}\widehat{p}_{k}+\bar{\bm{\psi}}_{k}\trans\widehat{\mathbf{p}}+\sigma_{k}^{2})
-\ln(\bar{\bm{\psi}}_{k}\trans\widehat{\mathbf{p}}+\sigma_{k}^{2}).
\label{eq:rate_def_1}
\end{equation}

Let $\widehat{\mathbf{p}}^{(\varkappa)}$ denote the feasible point at iteration $\varkappa$.  
Since the logarithm is concave, its first-order Taylor expansion provides a global underestimator.  
Linearizing the second (convex) logarithmic term in~\eqref{eq:rate_def_1} around $\widehat{\mathbf{p}}^{(\varkappa)}$ yields the concave lower bound
\begin{multline}
R_{k}(\widehat{\mathbf{p}})\geq
\ln(\psi_{k}\widehat{p}_{k}+\bar{\bm{\psi}}_{k}\trans\widehat{\mathbf{p}}+\sigma_{k}^{2})
-\ln(\bar{\bm{\psi}}_{k}\trans\widehat{\mathbf{p}}^{(\varkappa)}+\sigma_{k}^{2})\\
-\frac{\sum_{l=1}^{K}\bar{\psi}_{k,l}(\widehat{p}_{l}-\widehat{p}_{l}^{(\varkappa)})}
{\bar{\bm{\psi}}_{k}\trans\widehat{\mathbf{p}}^{(\varkappa)}+\sigma_{k}^{2}}
\triangleq R_{k}(\widehat{\mathbf{p}};\widehat{\mathbf{p}}^{(\varkappa)}).
\label{eq:rate_LB}
\end{multline}
Substituting~\eqref{eq:rate_LB} into~\eqref{eq:P3} leads to the convex surrogate formulation in~\eqref{eq:P4}. 
Hence,~\eqref{eq:P4} serves as a locally tight convex approximation of the original non-convex problem, which completes the proof.

\section{Proof of Theorem~\ref{thm:grad_theorem}}\label{sec:grad_proof}

Using $\nabla_{y_{n}}g_{k}(\mathbf{y},r_{k,\min},\varepsilon_{k})=-\nabla_{y_{n}}R_{k}(\mathbf{y})$,
$\nabla_{y_{0}}g_{k}(\mathbf{y},P_{\max},\varepsilon_{0})=\frac{1}{p_{\max}}\sum\nolimits_{k\in\mathcal{K}}p_{k}\nabla_{y_{n}}\tr(\widehat{\mathbf{\bm{\Sigma}}}_{k}(\mathbf{y}))$,
and~(\ref{eq:augObj}), a closed-form expression for $\nabla_{y_{n}}\mathscr{F}_{\bm{\nu},\rho}(\mathbf{y},\bm{\varepsilon})$
is given by~(\ref{eq:grad_closed}). On one hand, using~(\ref{eq:Rk}),
$\nabla_{y_{n}}R_{k}(\mathbf{y})$ can be written as 
\begin{equation}
\nabla_{y_{n}}R_{k}(\mathbf{y})=\bar{\tau}\frac{I_{k}(\mathbf{y})\nabla_{y_{n}}S_{k}(\mathbf{y})-S_{k}(\mathbf{y})\nabla_{y_{n}}I_{k}(\mathbf{y})}{(1+\gamma_{k}(\mathbf{y}))I_{k}^{2}(\mathbf{y})}.\label{eq:B-1}
\end{equation}
Following~(\ref{eq:closed_S}), we can obtain $\operatorname{d}(S_{k}(\mathbf{y}))$
as $\operatorname{d}(S_{k}(\mathbf{y}))=2p_{k}\tr(\widehat{\bm{\Sigma}}_{k}(\mathbf{y}))\tr(\mathbf{B}_{k}(\mathbf{y})\operatorname{d}\{\bm{\Sigma}_{k}(\mathbf{y})\})$, 
where $\mathbf{B}_{k}(\mathbf{y})\triangleq\mathbf{Q}_{k}(\mathbf{y})\bm{\Sigma}_{k}(\mathbf{y})-\mathbf{Q}_{k}(\mathbf{y})\bm{\Sigma}_{k}^{2}(\mathbf{y})\mathbf{Q}_{k}(\mathbf{y})+\bm{\Sigma}_{k}(\mathbf{y})\mathbf{Q}_{k}(\mathbf{y})$.
Hence, we have $\nabla_{y_{n}}S_{k}(\mathbf{y})=2A\mu_{k}p_{k}\tr(\widehat{\bm{\Sigma}}_{k}(\mathbf{y}))\tr(\mathbf{B}_{k}(\mathbf{y})\mathbf{O}_{n})$, 
where $\mathbf{O}_{n}\in\mathbb{R}^{N\times N}$ is an all-zero matrix,
except the $n$-th row. The $(n,m)$-th element (for $n\neq m$) of
$\mathbf{O}_{n}$ is given by 
\begin{equation}
[\mathbf{O}_{n}]_{n,m}=\bigg[\frac{\cos(2\pi d_{n,m}/\lambda)}{2\pi d_{n,m}/\lambda}-\frac{\sin(2\pi d_{n,m}/\lambda)}{2\pi d_{n,m}^{2}/\lambda}\bigg]\frac{y_{n}-y_{m}}{d_{n,m}},\label{eq:B-4}
\end{equation}
provided $n\neq m$ and $d_{n,m}\triangleq\|\mathbf{v}_{n}-\mathbf{v}_{m}\|$.
Next, following~(\ref{eq:closed_I}), we obtain $\operatorname{d}(I_{k}(\mathbf{y}))$
as $\operatorname{d}(I_{k}(\mathbf{y}))=\tr(\{\mathbf{C}(\mathbf{y})-2p_{k}\mathbf{D}_{k}(\mathbf{y})\}\operatorname{d}\{\bm{\Sigma}_{k}(\mathbf{y})\})
+\sum\nolimits_{\jmath\in\mathcal{K}}p_{j}\tr(\mathbf{G}_{k,\jmath}(\mathbf{y})\operatorname{d}\{\bm{\Sigma}_{\jmath}(\mathbf{y})\})$, 
where $\mathbf{C}(\mathbf{y})\triangleq\sum\nolimits_{j\in\mathcal{K}}p_{j}\widehat{\bm{\Sigma}}_{j}(\mathbf{y})$,
$\mathbf{D}_{k}(\mathbf{y})\triangleq\mathbf{Q}_{k}(\mathbf{y})\bm{\Sigma}_{k}(\mathbf{y})\widehat{\bm{\Sigma}}_{k}(\mathbf{y})-\mathbf{Q}_{k}(\mathbf{y})\bm{\Sigma}_{k}(\mathbf{y})\widehat{\bm{\Sigma}}_{k}(\mathbf{y})\bm{\Sigma}_{k}(\mathbf{y})\mathbf{Q}_{k}(\mathbf{y})+\widehat{\bm{\Sigma}}_{k}(\mathbf{y})\bm{\Sigma}_{k}(\mathbf{y})\mathbf{Q}_{k}(\mathbf{y})$,
and $\mathbf{G}_{k,j}(\mathbf{y})\triangleq\mathbf{Q}_{\jmath}(\mathbf{y})\bm{\Sigma}_{\jmath}(\mathbf{y})\bm{\Sigma}_{k}(\mathbf{y})-\mathbf{Q}_{\jmath}(\mathbf{y})\bm{\Sigma}_{\jmath}(\mathbf{y})\bm{\Sigma}_{k}(\mathbf{y})\bm{\Sigma}_{\jmath}(\mathbf{y})\mathbf{Q}_{\jmath}(\mathbf{y})+\bm{\Sigma}_{k}(\mathbf{y})\bm{\Sigma}_{\jmath}(\mathbf{y})\mathbf{Q}_{\jmath}(\mathbf{y})$.
Therefore, we have $\nabla_{y_{n}}I_{k}(\mathbf{y})=A\tr\big(\big\{\mu_{k}\{\mathbf{C}(\mathbf{y})-2p_{k}\mathbf{D}_{k}(\mathbf{y})\}
+\sum\nolimits_{\jmath\in\mathcal{K}}p_{j}\mu_{\jmath}\mathbf{G}_{k,\jmath}(\mathbf{y})\big\}\mathbf{O}_{n}\big)$.  
On the other hand, we have $\operatorname{d}\{\tr(\widehat{\bm{\Sigma}}_{k}(\mathbf{y}))\}=\tr(\mathbf{B}_{k}(\mathbf{y})\operatorname{d}\{\bm{\Sigma}_{k}(\mathbf{y})\})$.
Hence, we can write $\nabla_{y_{n}}\tr(\widehat{\bm{\Sigma}}_{k}(\mathbf{y}))=A\mu_{k}\tr(\mathbf{B}_{k}(\mathbf{y})\mathbf{O}_{n})$.  
This concludes the proof. 

\balance

\bibliographystyle{IEEEtran}
\bibliography{references}

\end{document}